\documentclass[
    aps,
    prl,
    twocolumn,
    superscriptaddress,letter
]{revtex4}

\usepackage{graphicx,psfrag}
\usepackage{amsmath,amsfonts,amssymb}
\usepackage{multirow}

\newcommand{\beq}{\begin{equation}}
\newcommand{\eeq}{\end{equation}}

\newcommand{\ket}[1]{| #1 \rangle}
\newcommand{\brahket}[3]{\langle #1 | #2 | #3 \rangle}

\newcommand{\mysum}[2]{\sum\limits_{#1}^{#2}}

\newcommand{\mybpar}[1]{\left( #1 \right)}

\newcommand{\beqa}{\begin{eqnarray}}
\newcommand{\eeqa}{\end{eqnarray}}

\newcommand{\Eqref}[1]{Eq.~(\ref{#1})}
\newcommand{\Figref}[1]{Fig.~\ref{#1}}

\begin{document}

\title{Unified description of inelastic propensity rules \\ for electron transport
through nanoscale junctions}
\author{Magnus Paulsson}
\email{magnus.paulsson@hik.se}
\affiliation{Division of Physics, School of Pure and Applied Natural Science,
University of Kalmar, 391 82 Kalmar, Sweden}
\affiliation{Department of Electronics, Toyama University, Gofuku,
Toyama,  930-8555, Japan}
\author{Thomas Frederiksen}
\affiliation{\mbox{Donostia International Physics Center (DIPC), Manuel de
    Landizabal Pasealekua 4,
20018 Donostia, Spain}}
\affiliation{CIC nanoGUNE Consolider, Mikeletegi Pasealekua 56, 20009 Donostia, 
Spain}
\author{Hiromu Ueba}
\affiliation{Department of Electronics, Toyama University, Gofuku,
Toyama,  930-8555, Japan}
\author{Nicol\'as Lorente}
\affiliation{Centro de Investigaci\'on en Nanociencia y Nanotecnolog\'ia, 
CSIC-ICN, Campus de la Universitat Aut\`onoma de Barcelona, 08193 Bellaterra,
Spain}
\affiliation{Department of Electronics, Toyama University, Gofuku,
Toyama,  930-8555, Japan}
\author{Mads Brandbyge}
\affiliation{DTU Nanotech, NanoDTU, Technical University of Denmark, 2800 Lyngby, Denmark}

\date{\today}

\begin{abstract}
We present a method to  analyze the results of
first-principles based calculations of electronic currents
including inelastic electron-phonon effects. 
This method allows us to determine the electronic and vibrational
symmeties in play, and hence to obtain the so-called {\em propensity
rules} for the studied systems. 
We show that only a few scattering states --- namely those belonging 
to the most transmitting eigenchannels --- need to be considered for 
a complete description of the electron transport.
We apply the method on first-principles 
calculations of four different
systems and obtain the propensity rules in each case.
\end{abstract}

\maketitle

%\section{Introduction}

Electronic transport through atomic-size junctions is of immense scientific and
technological interest.  
The importance of inelastic effects
in electronic currents have been revealed in several ground-breaking
experiments leading to the
detection and identification of single molecules~\cite{Stipe}, 
chemical reactions~\cite{chemistry1,chemistry2},
the detection of vibrations in atomic wires~\cite{Agrait2002}, 
the detection of inelastic effects by fluorescence~\cite{Qiu}, 
the modification of electron transport in nanotubes~\cite{LeRoy}, 
the molecular motion induced by electronic currents ~\cite{Lastapis},
and the hydrogen detection in atomic wires~\cite{Kiguchi}, just
to cite a few examples. Of particular importance due
to its spreading use is the case of vibrational spectroscopy where
the conductance changes due to phonon emission is measured 
\cite{Smit2002,Kushmerick2004,Troisi2007}.  
This is often referred to as point contact spectroscopy or inelastic
electron tunneling spectroscopy (IETS)~\cite{Stipe}, 
%Recently, IETS has been used 
%to study mechanical properties of atomic wires
% \cite{Agrait2002} 
%and to identify molecules in nanocontacts \cite{Djukic2006,Kushmerick2004,Troisi2007}. 
However,  experiments alone are not able to give direct insight into
the fundamental question on how the detailed atomic structure 
correlate with the electrical transport
properties. 
There is experimental evidence
of {\em approximate} selection rules ({\em propensity rules}~\cite{Troisi2006})
 such that
only a small number out of the many
possible vibrational modes give an
inelastic signal. 
These propensity rules yield clues to the geometric and 
electronic structure of the junctions. 
It is therefore of fundamental interest to compare the experimental results
with first-principles calculations.

Existing calculations of inelastic effects in electron transport
have been developed either for particular cases~\cite{Lorente2001,Troisi2006,Gagliardi2007} or for 
simplified (one-level)
models~\cite{Mii2003,Galperin2008}. First-principles methods capable of
treating both weak and strong coupling to the electrodes has also been
developed \cite{Paulsson2005,Paulsson2006,Frederiksen2007}. However, the
results of such detailed calculations involve many electronic states and
vibrational modes. An advanced analysis is therefore needed in order to
provide insight into the propensity rules.  

In this paper we propose a method for analysis of the inelastic transport
based on just a few {\emph{selected}} electronic scattering states, namely
those belonging to the most transmitting eigenchannels at the Fermi energy
($\varepsilon_f$) \cite{Paulsson2007}.
%such an
%analysis focusing on the small number of scattering states which can propagate
%into an atomic sized junction and conduct current, i.e., transmission 
%eigenchannels \cite{Paulsson2007}. 
These scattering states typically have the 
largest amplitude inside the junction and thus account for the majority of the
electron-phonon (e-ph) scattering. To
illustrate our method of analysis and to develop an understanding of the
propensity rules we consider four cases: (i)
atomic gold-wires, and (ii) molecular junctions, as well as scanning tunneling microscope (STM)
setups in the (iii) resonant, and (iv) non-resonant limits. 
The propensity rules can in these cases be understood from
e-ph induced transitions between scattering states of a few eigenchannels.

%\section{Method}

Inelastic scattering of electrons in a device under bias can be  modeled using
nonequilibrium Green's functions (NEGF) 
\cite{Mii2003,Frederiksen2007,Galperin2008}.  In particular, the
lowest order expansion 
(LOE) of the NEGF equations provides a tractable description of
phonon scattering in first-principles
calculations \cite{Paulsson2005}. This approximation assumes a weak
e-ph coupling ($\mathbf{M}$) and that the  electronic structure
changes slowly over a phonon energy ($\hbar \omega$). It is therefore not
applicable to strong e-ph coupling. In the zero-temperature limit, 
the conductance is  
\beq
 G^\mathrm{LOE} =   G_0 \tau  +
  \mysum{\lambda}{} 
  e \gamma_\lambda^{\mathrm{LOE}} \theta( \left| eV \right| - \hbar \omega_\lambda)+
  {\cal G}^\mathrm{Asym}_\lambda  \label{eq.current1} ,
\eeq
which can be divided into the Landauer term, with the transmission
$\tau$ (at $\varepsilon_f$) times the conductance quantum
$G_0$, and inelastic corrections in the conductance
from each vibrational mode.  In this formulation we have separated the
inelastic contribution in a symmetric term, with respect to bias, from phonon
absorption and emission processes, and an asymmetric term ${\cal
  G}^\mathrm{Asym}_\lambda$. ${\cal G}^\mathrm{Asym}_\lambda$ is small in the
cases studied here since it is (i) strictly zero for symmetric junctions and
(ii) negligible in the STM configuration when very close to or very far from
resonance \cite{Mathematica2007,Paulsson2005}. For these reasons we 
ignore ${\cal  G}^\mathrm{Asym}_\lambda$ in the following discussion.
Each mode thus gives a step 
$\theta(|eV|-\hbar\omega_\lambda)$ in the conductance at the phonon energy. 
The magnitude and sign of this
step is given by the scattering rate per excess bias 
\begin{eqnarray}
\gamma_\lambda^{\mathrm{LOE}} &=&  \frac e {\pi \hbar} \mathrm{Tr}\big[
  \mathbf{G}^\dag \mathbf{\Gamma}_\mathrm{L} \mathbf{G} \big\{ \mathbf{M}_\lambda \mathbf{G}
  \mathbf{\Gamma}_\mathrm{R} \mathbf{G}^\dag
  \mathbf{M}_\lambda \nonumber\\
&& +\frac{i}{2}\big( \mathbf{\Gamma}_\mathrm{R} \mathbf{G}^\dag 
  \mathbf{M}_\lambda \mathbf{A} \mathbf{M}_\lambda - \mathrm{h.c.}\big)
  \big\}\big], \label{eq:LOE-rate}
\end{eqnarray}
where $\mathbf{G}$ is the retarded Green's function at $\varepsilon_f$, 
$\mathbf{A}=i(\mathbf{G}^{\phantom\dagger}-\mathbf{G}^{\dagger})$ the spectral function, 
and $\mathbf{\Gamma}_{\mathrm{L,R}}$ the couplings to the leads. 

\begin{figure}[t]
\psfrag{alpha}[c][c]{$\alpha=\Gamma_{\mathrm{R}}/\Gamma_{\mathrm{L}}$} 
\psfrag{tau}[c][c]{Transmission $\tau$}
\psfrag{g1}[c][c]{$\Gamma_{\mathrm{L}}$}
\psfrag{g2}[c][c]{$\Gamma_{\mathrm{R}}$}
\psfrag{e0}[c][c]{$\epsilon_0$}
\psfrag{Molwire}[c][c]{Molecular wires}
\psfrag{Auwire}[c][c]{Atomic wires}
\includegraphics[width=1.0 \columnwidth,clip=]{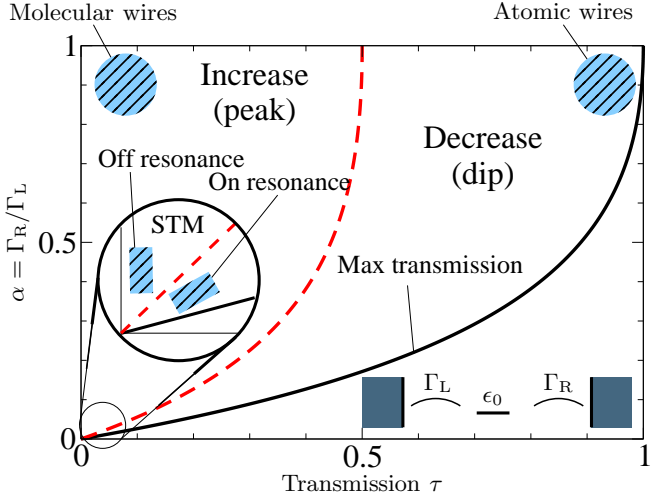}
\caption{Phase diagram for a one-level model (inset) illustrating
the sign of the conductance change at the onset of phonon
emission. At a given asymmetry factor $\alpha$ the elastic
transmission $\tau$ has an upper bound $\tau_\textrm{max}$  (black line), 
and the inelastic conductance change undergoes a sign change at 
$\tau_\textrm{crossover} = \tau_{\mathrm{max}}/2$ (red dashed line). 
The one-level model is available online \cite{Mathematica2007}.} 
\label{fig.phase}
\end{figure}

It is instructive to use a one-level model to get an 
understanding of the sign of the conductance step. 
The one-level model couples a single electronic level to two electronic leads
and a localized vibration \cite{Paulsson2005,Mathematica2007}.
A phase diagram is plotted in Fig.~\ref{fig.phase} 
for the ratio of the coupling to the two leads
$\alpha=\Gamma_{\mathrm{R}}/\Gamma_{\mathrm{L}}$
and the transmission $\tau$ at $\varepsilon_f$. For this model the 
maximal transmission is 
$\tau_{\mathrm{max}}={4 \alpha}/{(1+\alpha)^2}$ corresponding
to the on-resonance case. 
The crossover from a decrease to an increase 
in the conductance is given by the \emph{1/2 rule} \cite{tal2008,Paulsson2005,Vega2006},
i.e., at $\tau_{\mathrm{crossover}}=\tau_{\mathrm{max}}/2$. 
%\footnote{The asymmetric term mentioned in conjunction with 
%  \Eqref{eq.current1} has a maximum at 
%  the crossover \cite{Mathematica2007}. In addition, at the crossover
%  the electronic structure is non-constant as a function of energy. 
%  For these reasons the 1/2 rule should be considered an approximate rule.}.
Back-scattering dominates the high-transmission 
system leading to a decrease of the conductance while 
forward-scattering leads to an increase in the low-transmitting case
\cite{Ueba2007}.
%\footnote{For a low-transmission systems on resonance, 
%  the elastic part of the conductance decreases more than the increase from
%  the inelastic part .}. 

The scattering rate $\gamma^{\mathrm{LOE}}$ may be interpreted as a competition between an
inelastic process, first term of \Eqref{eq:LOE-rate}, that increases the current 
and an elastic correction, second term of \Eqref{eq:LOE-rate}, that
decreases the current \cite{Mii2003,Lorente2001,Ueba2007}. 
In the case of low transmission, where only the inelastic term needs 
to be considered, Troisi {\it et al.} \cite{Troisi2006} and
Gagliardi {\it et al.}  \cite{Gagliardi2007} have extensively discussed the
propensity rules.
% using the abstract notion of Green's functions and
% eigenvectors of $\mathbf \Gamma_\textrm{L,R}$, respectively.

Instead of trying to understand the complex issues of the competition between
elastic and inelastic parts of the conductance, we have found that the phonon
emission rate provides a simple way to obtain the IETS propensity rules. 
In the LOE approximation the power deposited into the phonon system is 
given by  \cite{Paulsson2005}   
\beqa
P^\mathrm{LOE}&=& \mysum{\lambda}{}
  \frac{\mybpar{\hbar \omega_\lambda}^2}{\pi \hbar}
    \big\{ n_\textrm{B}(\hbar \omega_\lambda)  -n_\lambda \big\}
    \, \mathrm{Tr}\left[ \mathbf{M}_\lambda \mathbf{A} \mathbf{M}_\lambda \mathbf{A} \right]
   \nonumber \\
&& + 
    {\hbar \omega_\lambda} \, \gamma_\lambda^{\mathrm{FGR}}
   \left\{ \begin{array}{*{2}{l}} 0 & ; \left| eV \right| < \hbar \omega_\lambda\\\left|
  V\right|-\frac{\hbar \omega_\lambda}{e}  & ; \left| eV \right| >
  \hbar \omega_\lambda \end{array} 
    \right. .
  \label{eq.power1} 
\eeqa
The first term describes electron-hole damping of the
vibrations that drives the actual occupation $n_\lambda$ towards
the Bose-Einstein equilibrium value $n_\textrm{B}$.
The second term describes 
the heating of the phonon system in terms of the emission rate  \cite{Paulsson2005}   
\beq
    \gamma_\lambda^{\mathrm{FGR}}=\frac{e}{\pi \hbar}
      \mathrm{Tr}\left[ \mathbf{A}_\mathrm{L} \mathbf{M}_\lambda
      \mathbf{A}_\mathrm{R} \mathbf{M}_\lambda \right] =
    \frac{4 \pi e}{\hbar}  \sum\limits_{l,r}
    \left| \brahket{\Psi_l}{\mathbf{M}_\lambda}{\Psi_r}\right|^2, \label{eq.FGR}
\eeq
where $\mathbf{A}_{\mathrm{L,R}}=\mathbf{G}\mathbf{\Gamma}_{\mathrm{L,R}}\mathbf{G}^\dagger$ 
are the partial spectral functions from the two leads. To provide physical
insights we rewrite the trace in terms of a complete set of scattering states
$\ket{\Psi_{l,r}}$ from the left (right) leads. This gives Eq.~(\ref{eq.FGR}) in
the form of the physically transparent Fermi's golden rule (FGR). It is
advantageous to choose the basis as eigenchannels \cite{Paulsson2007}, i.e.,
the scattering states belonging to the largest transmission. Since the e-ph
coupling is essentially local in space, it is sufficient to evaluate
Eq.~(\ref{eq.FGR}) using only a few of the most transmitting eigenchannels while
the reflected scattering states can be ignored. For the examples described
below, only one to three scattering states are needed to account for over 90\%
of the phonon scattering.

\begin{figure}[t]
\psfrag{IETS}[c][c]{$\mbox{d}^2 I/\mbox{d}V^2/(\mbox{d} I/\mbox{d}V)\,\,(1/V)$}
\psfrag{dEf123456789}[c][c]{$\Delta \varepsilon_f=-0.6$ eV}
\psfrag{(a)OPE/Au}[l][l]{(a) OPE/Au}
\psfrag{(b)Au-wire}[r][r]{(b) Au-wire}
\psfrag{(c)CO/Cu}[l][l]{(d) CO/Cu}
\psfrag{(d)O2/Ag}[r][r]{(c) O$_2$/Ag}
\psfrag{Theory}[l][l]{Theory}
\psfrag{Exp.}[l][l]{Exp.}
\psfrag{Bias[mV]}[c][c]{Bias (mV)}

\includegraphics[width=\columnwidth,clip=]{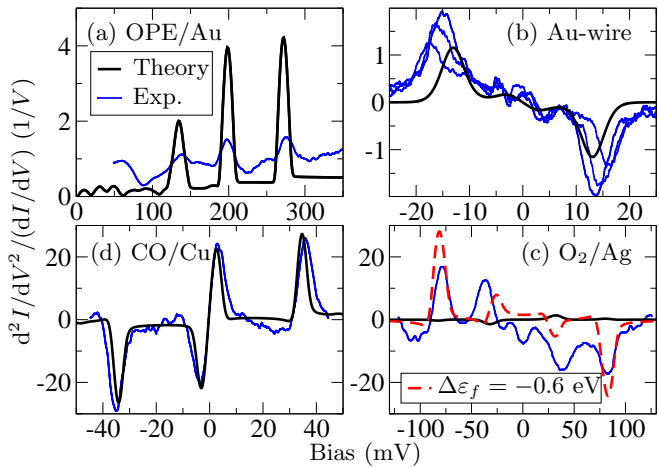}
\caption{Calculated (black lines) and experimental
 (blue lines) IETS representing the 
corners of the phase diagram in Fig.~\ref{fig.phase}: 
(a) OPE molecule with Au(111) leads,
(b) Au chain connected to Au(100) leads,
(c) O$_2$ molecule on Ag(110), and
(d) CO molecule on Cu(111).
%The IETS is shown with the Fermi energy as calculated by DFT (solid black) and 
%manually shifted down by 0.6 eV (dashed red) for the O$_2$/Ag system as 
%discussed in the text. 
In case (c) the Fermi energy ($\varepsilon_f$) as been 
shifted manually to match the experiment
(dashed red line).
The experimental data originates from
Refs.~ \cite{Agrait2002,Kushmerick2004,Heinrich2002,Hahn2000}. 
For the STM configurations (c) and (d), the calculated IETS is compared with 
a rescaled $\mbox{d}^2 I/\mbox{d}V^2$.}
\label{fig.results}
\end{figure}

%\section{Results}
To illustrate how the phonon emission rate leads to the IETS propensity rules
we have performed calculations on four experimentally realized systems which
we believe corresponds to the four 
'corners' of the phase diagram, see Fig.~\ref{fig.phase} and \ref{fig.results}. 
The calculations were performed within density functional theory (DFT) 
\footnote{A discussion of the DFT related issues with the position of
the $\pi^*$ resonance of $\mathrm{O}_2$/Ag(111) \cite{Olsson2003} and 
the CO adsorption on Cu(111) are outside the scope of 
this work. \label{fot.dft}}
using our extension of TranSIESTA
%the SIESTA/TranSIESTA 
%computer codes, and the geometry, phonon energies, e-ph
%coupling, and electronic structure including coupling to infinite
%leads were calculated
% in the Gamma-point approximation with the
%generalized gradient approximation Perdew-Burke-Ernzerhof functional
%using localized basis sets
as described in Ref.~ \cite{Frederiksen2007}. Broadening by temperature 
and lock-in modulation $V_\mathrm{rms}$ were included.

The IETS, defined as $\mbox{d}^2
I/\mbox{d}V^2/(\mbox{d} I/\mbox{d}V)$, is shown  
in Fig.~\ref{fig.results} for:
(a) Symmetric low-transmission case, an oligo-phenyl-ethylene (OPE)
molecule symmetrically thiol-bonded to 
the hollow position on Au(111) leads.
The temperature and modulation voltage used in the calculation were
$T=4.2$ K and $V_\mathrm{rms}=8$ meV.  
As we have described previously  \cite{Paulsson2006}, the calculated IETS compare qualitatively 
with measurements  \cite{Kushmerick2004}.
(b) Symmetric high-transmission case, a 7-atom Au chain connected to
Au(100) leads, $T=4.2$ K, $V_\mathrm{rms}=1$ meV. There is quantitative agreement
with experiments  \cite{Agrait2002,Frederiksen2007}. 
(c) On-resonance STM configuration with an O$_2$ molecule on Ag(110)
surface displaying a decrease in the conductance upon
phonon emission  \cite{Hahn2000}. The STM tip is modeled by a single Ag atom on a
Ag(110) lead laterally displaced by $1.6$ \AA ~ corresponding to the experimental
situation \cite{Hahn2000}, 
$T=13$ K, $V_\mathrm{rms}=7$ meV.  
The IETS for O$_2$/Ag(110) is shown both for the self-consistent DFT calculation
(solid black) as well as with $\varepsilon_f$ 
shifted manually by $-0.6$ eV with respect to the DFT result
(dashed red, low bias conductance 750 nA/V). 
(d) Off-resonance STM configuration with a CO molecule adsorbed on a
Cu(111) surface  \cite{Heinrich2002}. 
The STM tip is modeled by a single Cu atom positioned on a Cu(111)
lead, $T=5$ K, $V_\mathrm{rms}=2$ meV.  
%The differences are discussed later on. 

\begin{figure}[t]
\psfrag{dI}[c][c]{$\mbox{d}^2 I/\mbox{d}V^2/(\mbox{d}
  I/\mbox{d}V)\,\,(1/V)$}
\psfrag{rate1}[c][c]{$\gamma  \, \left(10^{11}/(sV)\right)$}
\psfrag{rate2}[c][c]{$-\gamma  \, \left(10^{12}/(sV)\right)$}
\psfrag{IETS}[l][l]{IETS}
\psfrag{FGRFG}[l][l]{$\left|\gamma_{\mathrm{FGR}}\right|$}
\psfrag{LOE}[l][l]{$\left|\gamma_{\mathrm{LOE}}\right|$}
\psfrag{BIAS}[c][c]{Bias, $\hbar \omega \, \left(\mathrm{meV}\right)$}
\psfrag{OPE}[l][l]{(a) OPE/Au}
\psfrag{Au}[l][l]{(b) Au-wire}

\includegraphics[width=1.0\columnwidth,clip=]{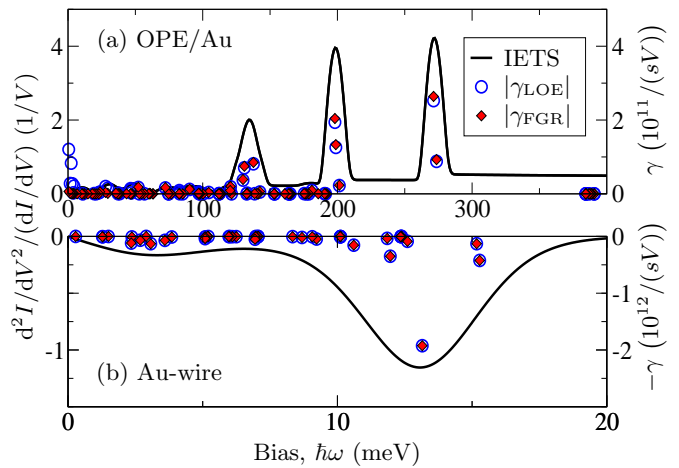}
\caption{Comparison between the FGR scattering rate, calculated using
  only the single-most transmitting eigenchannel, with the full LOE
  rate for (a) OPE molecule with Au(111) leads (zero-bias conductance
  of $1.0 \times 10^{13}/(\mathrm{sV})$), and (b) 7-atom Au wire ($4.8
  \times 10^{14}/(\mathrm{sV})$). The sign of the FGR rate was chosen
  to agree with the LOE rate.}
\label{fig.results2}
\end{figure}

We find that the size of the conductance drop given by 
$\gamma^\mathrm{LOE}$ is well approximated with the phonon emission rate $\gamma^\mathrm{FGR}$. 
%(repetition) The phonon emission rate \Eqref{eq.FGR} thus gives to a good approximation the  
%size of the conductance step $\gamma^\mathrm{LOE}$. 
For the OPE and Au-wire cases, the first eigenchannel gives the majority of the phonon emission rate, see Fig.~\ref{fig.results2}. 
%Thus, the propensity rules are readily understood from the eigenchannels involved, i.e., 
%the eigenchannels select the IETS signals (except for the sign of the
%conductance change).
%although the sign of the conductance change has to be inferred from 
%the transmissions of the involved eigenchannels.
In this case the 
propensity rules follows from the symmetry of the 
scattering states which are similar 
to the Bloch states of the corresponding periodic systems \cite{Paulsson2007}. 
To illustrate this understanding of the IETS we have published
Mathematica notebooks showing the propensity rules for tight-binding models \cite{Mathematica2007}. 
Before comparing the theoretical STM-IETS to the experimental, we note
that the typical tip-sample distances of STM are much larger than what
is feasible computationally with a localized basis set. In practice, we work
with a significantly smaller tip-sample distances than in the experimental
situation, while still being in the tunneling regime.

%\emph{TF: The following discussion is rather long}
%Before comparing the theoretical IETS to the experimental results for
%the two cases we note that 
%the tip-sample distance in a STM measurement is determined by the set current. 
%This low bias conductance is, in general, a very small tunneling conductance of the order of nS.
%Our use of localized basis sets in the DFT calculations limits the range of the electronic 
%wave-function. Tunneling between atoms is therefore only possible
%between atoms with overlapping  
%basis functions. In practice this means that we carry out the calculations with a smaller 
%tip-sample distance and use a basis set with a longer than commonly used range. 
%To compare the experimental IETS for CO on Cu(111)
%surface and O$_2$ on Ag(110), we have therefore rescaled the
%experimental IETS to coincide with the height of the calculated IETS
%peaks in Fig.~\ref{fig.results} (c) and (d). 

Results for the O$_2$ on Ag(110) system
is shown in Fig.~\ref{fig.results}(c). In this case
our first-principles calculation fails to describe the conductance 
decreases observed experimentally \footnotemark[30]. 
The DFT-based energy spectrum does not have a molecular resonance at 
$\varepsilon_f$, c.f., the on-resonance
case of \Figref{fig.phase}. 
However, DFT predicts a $\pi$* resonance 0.6 eV below $\varepsilon_f$ 
\cite{Olsson2003}. By manually adjusting $\varepsilon_f$ 
to this resonance we manage to capture some of the 
qualitative features of the IETS. Although this result is 
suggestive, we have no conclusive evidence that the experimental 
decrease is caused by this mechanism.  
 
%We can therefore only
%conclude that although it is possible that the $\pi$* resonance is responsible for the
%conductance drop, conclusive evidence is missing and we only include this preliminary
%result to exemplify the lower right hand corner of Fig.~\ref{fig.phase}. 

\begin{figure}[t]
\psfrag{EC1}[c][c]{Eigenchannel 1}
\psfrag{EC2}[c][c]{Eigenchannel 2}
\psfrag{FR}[l][l]{FR(x)}
\psfrag{FT}[l][l]{FT(x)}
\psfrag{hwFR}[l][l]{34 meV}
\psfrag{hwFT}[l][l]{3 meV}
\psfrag{t1}[l][l]{t1}
\psfrag{t2}[l][l]{t2}
\psfrag{s1}[l][l]{s1}
\psfrag{s2}[l][l]{s2}
\psfrag{tip}[c][c]{tip}
\psfrag{substrate}[c][c]{substrate}
\psfrag{a}[c][c]{(a)}
\psfrag{b}[c][c]{(b)}
\psfrag{x}[c][c]{$\hat x$}
\psfrag{y}[c][c]{$\hat y$}
\psfrag{z}[c][c]{$\hat z$}

\includegraphics[width=1.0\columnwidth,clip=]{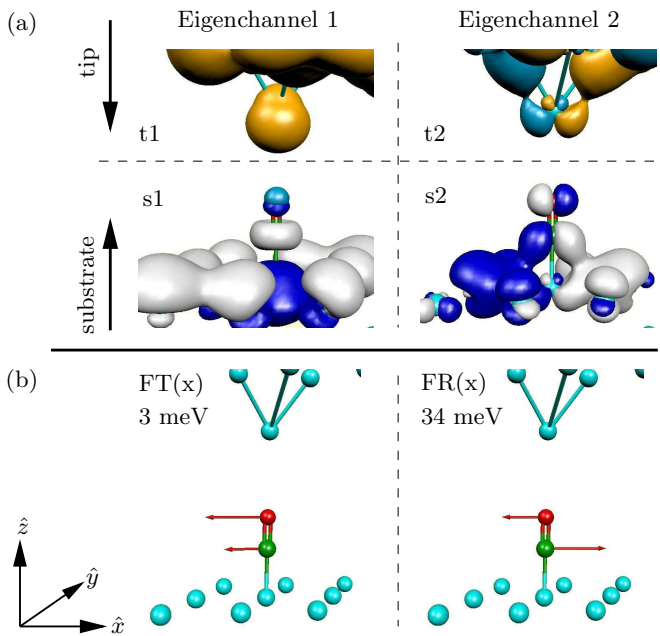}
\caption{Eigenchannels and phonon modes for CO on Cu(111). (a)
  Scattering states associated with
eigenchannel 1 (left column, $\tau_1=1.9\times 10^{-3}$) and 
eigenchannel 2 (right column, $\tau_2=0.4\times 10^{-3}$). 
The top (bottom) row of scattering states originate from the tip
(substrate) side.
Eigenchannel 3 (not shown,  $\tau_3=0.4\times 10^{-3}$) closely 
resemble those of eigenchannel 2 rotated by 90 degrees.
(b) FR(x) and FT(x) vibrational modes (degenerate with FR(y) and FT(y)).
} \label{fig.EC}
\end{figure}

The results for CO/Cu(111) \footnotemark[30] is shown in Fig.~\ref{fig.results}(d)
and \Figref{fig.EC}. The height of the IETS peaks are qualitatively 
captured, we predict that 13 $\%$ of the transmitted 
electrons will emit a frustrated rotation phonon compared to $8 \%$ 
found experimentally for CO/Cu(001) \cite{Lauhon1999,Persson2004}. 
The eigenchannels necessary to calculate the transition rates 
$\gamma^\mathrm{FGR}$ are shown in Fig.~\ref{fig.EC}(a).
% where each
%eigenchannel has a scattering state incident from the tip as well as from the  
%substrate lead.
%The corresponding scattering states
%originating from the tip side ($t1, t2$) is plotted
%in the top part while those originating from the substrate side ($s1, s2$)
%is plotted in the lower part.
We note that the primary eigenchannel is of $\sigma$-type, i.e., rotationally symmetric
around the tip-molecule direction, while the secondary and tertiary
are of $\pi$-type. Comparing the scattering states to the phonon modes,
Fig.~\ref{fig.EC}(b), the propensity rules naturally follows from the
symmetry. 
The frustrated translation (FT$_{x,y}$) 
and frustrated rotation (FR$_{x,y}$) modes (in the surface $xy$-plane) 
can only scatter between a $\pi$- and a $\sigma$-type eigenchannel,
because the modes have $\pi$-character with respect to the transport
direction, i.e., $s2\leftrightarrow t1$ and $s3\leftrightarrow t1$. 
Note that the $s1 \leftrightarrow t2,3$ transitions are less important due to
the weight of the scattering states (Fig.~\ref{fig.EC}). 
Similarly, the $\sigma$-type FT$_z$ and CO stretch modes 
only scatter between 
eigenchannels of the same type, e.g., $s1 \leftrightarrow t1$. 
Table \ref{tab.modes} confirms these observations and gives a 
quantitative account of the e-ph scattering in the CO/Cu(111) system. 
It shows that $\gamma^\mathrm{FGR}$ closely approximates the $\gamma^\mathrm{LOE}$ and that 
the main part 
%of $\gamma^\mathrm{FGR}$ 
is given by the first three eigenchannels. 
%TF: I think the following has more or less been described:
%Furthermore, the fact that the frustrated rotation and translation
%modes scatter predominantly within from the second and third eigenchannel from the
%substrate side to the first eigenchannel on the tip side is due to the size of
%the weight of the eigenchannel wavefunctions in the junction region, see
%Fig.~\ref{fig.EC}(a).

%TF: This is an interesting comment but due to length limitations...
%It is interesting to note that the different eigenchannels may have different
%exponential decay into the tunnel gap. It is therefore possible that the
%relative scattering amplitudes may depend on the actual tip-sample distance,
%and one can speculate whether this can be used to obtain further information 
%from experiments.

\begin{table}[t]
\begin{tabular}{rcccl}
$\hbar \omega$ (meV) & 
FGR (LOE) & $\mathrm{subs.} \leftrightarrow
\mathrm{tip}$  & \% & Mode\\
\hline
\hline
 236 & 0.9 (0.8) &$s_i\leftrightarrow t_i, i=1,2,3$& 100  & CO stretch  \\
 48 & 0.3 (0.3) & $s_i\leftrightarrow t_i, i=2,3$ & 95 &  FT(z)\\
 35 & 8.2 (8.0) & s3 $\leftrightarrow$ t1 & 95 & FR(y) \\
 34 & 8.3 (8.1) & s2 $\leftrightarrow$ t1 & 95 & FR(x) \\
 3 & 5.9 (5.8) & s3$\leftrightarrow$ t1 & 92 & FT(y)\\
 3 & 6.1 (6.0) & s2$\leftrightarrow$ t1 & 92 & FT(x)\\
 \hline
\end{tabular}
\caption{Vibrational modes and inelastic scattering rates for CO on Cu(111). 
The FGR rate \Eqref{eq.FGR} (using all eigenchannels) and the LOE rate
\Eqref{eq.current1} are given in units of $10^{10}(sV)^{-1}$ (elastic
transmission $=130\times10^{10}\,(sV)^{-1}$).
The dominating transitions between the eigenchannel
scattering states
%responsible for the main scattering amplitude  
are indicated along with the fraction in which these processes contribute
to the total FGR rate. 
% TF: Do we strictly need the following information?
%The calculation was performed with a Cu atom on top of a Cu(111)
%surface to model the STM tip with a  
%zero-bias conductance of 
%$2.7\times10^{-3} \, G_0 = 208\,
%\mathrm{nA/V}=130\times10^{10}\,(sV)^{-1}$
}
\label{tab.modes}
\end{table}

%\section{Conclusions}

We have presented a method to analyze inelastic e-ph
scattering in terms of eigenchannel scattering states. 
The main advance in the context of first-principles transport
calculations is to bring the description into a natural framework where
underlying symmetries of the propensity rules can be
understood. Through four different examples 
corresponding to different transport regimes, we demonstrated 
that the eigenchannel analysis addresses the propensity rules 
in a unified way.

%Finally we note that the presented tools for analyzing propensity
%rules are not inherently bound to DFT, but to any scheme ... Not limited to DFT?

%For first-principles calculations the simplification offered by only having to include a few
%electronic channels and the use of a FGR approach allows the
%intuitive understanding of inelastic propensity rules. In addition, the DFT
%based calculations presented here shows the success of DFT to
%describe some properties of electron transport through atomic- and
%molecular-junctions. This success is remarkable since we use the Kohm-Sham
%orbitals to describe the electrons and perhaps more effort should be aimed at
%understanding systems where the method is less successful, i.e., the O$_2$
%molecule on Ag surface.

%\bibliographystyle{apsrev}
%\bibliography{FGRbib}

\end{document}